\documentclass[review]{elsarticle}

\usepackage{xcolor}

\usepackage[labelsep=endash]{caption}

\usepackage[colorlinks=true,
            linkcolor=red,
            citecolor=blue,
            urlcolor=red
            ]{hyperref}

\usepackage{caption}
\usepackage{graphicx}

\usepackage{subfigure}
\usepackage{subcaption}
\usepackage{multirow}
\usepackage{gensymb}
\usepackage{tabularx}
\usepackage{booktabs}
\usepackage{amsmath}
\usepackage{accents}
\usepackage[export]{adjustbox}
\usepackage[margin=3.3cm]{geometry}
\usepackage{setspace}
\usepackage[pagewise]{lineno}
\usepackage{lipsum}
\makeatletter

\journal{×××××××××××××××××××~~~~}

\begin{document}
\nolinenumbers
\begin{frontmatter}

\title{A 3D virtual geographic environment for flood representation towards risk communication}





\author[mymainaddress,mysecondaryaddress,myfourthaddress]{Weilian Li}

\author[mymainaddress]{Jun Zhu\corref{mycorrespondingauthor}}\cortext[mycorrespondingauthor]{Corresponding author}
\ead{zhujun@swjtu.edu.cn}

\author[mythirdaddress]{Saied Pirasteh}

\author[mymainaddress]{Qing Zhu}

\author[mymainaddress]{Yukun Guo}

\author[mymainaddress]{Lan Luo}

\author[mysecondaryaddress]{Youness Dehbi}

\address[mymainaddress]{Faculty of Geosiences and Engineering, Southwest Jiaotong University,
Chengdu 611756, China}
\address[mysecondaryaddress]{Computational Methods Lab, HafenCity University Hamburg, Hamburg 20457, Germany}
\address[mythirdaddress]{Institute of Artificial Intelligence, Shaoxing University, Shaoxing 312000, China}
\address[myfourthaddress]{Guangdong–Hong Kong-Macau Joint Laboratory for Smart Cities, Shenzhen 518060, China}

\begin{abstract}
Risk communication seeks to develop a shared understanding of disaster among stakeholders, thereby amplifying public awareness and empowering them to respond more effectively to emergencies. However, existing studies have overemphasized specialized numerical modelling, making the professional output challenging to understand and use by non-research stakeholders. In this context, this article proposes a 3D virtual geographic environment for flood representation towards risk communication, which integrates flood modelling, parallel computation, and 3D representation in a pipeline. Finally, a section of the Rhine River in Bonn, Germany, is selected for experiment analysis. The experimental results show that the proposed approach is capable of flood modelling and 3D representation within a few hours, the parallel speedup ratio reached 6.45. The intuitive flood scene with 3D city models is beneficial for promoting flood risk communication and is particularly helpful for participants without direct experience of floods to understand its spatiotemporal process. It also can be embedded in the Geospatial Infrastructure Management Ecosystem (GeoIME) cloud application for intelligent flood systems.
\end{abstract}


\end{frontmatter}

  \section{Introduction}

As global climate change continues to exacerbate the extreme weather, floods have become the most frequent and destructive natural disaster leading to severe social and economic impacts \citep{costabile2021terrestrial,li2022investigations,habibi2023novel,habibi2023hybrid,khorrami2023statistical,guo2023dynamic}. In July 2021, summer storms and severe weather resulted in significant rainfall. The catastrophic floods severely affected some European countries (e.g., Germany, Belgium, Netherlands.), and caused more than 700 injuries and almost 200 deaths \citep{li2023social}. The economic losses amount to approximately 35.3 billion euros\footnote{\url{https://us.milliman.com/-/media/milliman/pdfs/2022-articles/3-28-22_europe-extreme-weather-report.ashx}}. At almost the same time, intensive floods also occurred in China and the United States, this is a hint that the flood impacts are expected to increase, and become a global issue in the near future \citep{dottori2018increased,berndtsson2019drivers,costabile2021terrestrial, luo2022increasing}.

Besides, countries and emergency management have struggled to enhance flood mitigation and loss reduction worldwide \citep{pirasteh2019geospatial}. In the last decade, the focus of flood mitigation has shifted from flood protection to risk management \citep{costabile2021terrestrial,li2021augmented,zhu2024knowledge}. As an integral part of flood risk management, risk communication is an interactive process of information exchange, which aims to share flood information among stakeholders, convince the population at risk to be prepared for an emergency, and improve risk awareness \citep{hagemeier2009evaluation,dransch2010contribution,macchione2019moving,zhu2024knowledge,zhu2024flood}.

In the early stages, flood risk information is generally communicated to the people by radio, TV, and community staff door-knocking in some circumstances. Subsequently, geoinformatics and hydrological modelling are introduced into flood risk management. The most representative product is the flood risk/hazard map, which overlays 2D flood modelling output with resource maps to improve government preparedness and individual response to floods \citep{ntajal2017flood,seipel2017color,henstra2019communicating,taylor2020messy,costabile2021terrestrial,li2022investigations}. The flood map is straightforward, but it is not intuitive due to poor topographic auxiliary information compared with the 3D representation \citep{yang2016gis,zhang2020efficient}. 

More precisely, 3D representation enables the effortless understanding of complicated flood phenomena and presents its process without temporal and spatial impediments \citep{yang2016gis,herman2017flood}, which has been proven to enhance communication among various stakeholders and improve the effectiveness of decision-making processes \citep{wang2006visualizing,bandrova2012three,qiu2017integrated,zhang2020efficient,costabile2021terrestrial,li2024visual}.  For example, flood representation in 3D enhances awareness and minimizes losses associated with flooding. It offers a realistic visualization of flood scenarios, clearly understanding potential risks. This immersive experience aids in better preparation and community engagement \citep{li2024visual}. Decision-makers can use it to evaluate flood management strategies, while emergency responders can train using simulated scenarios. Additionally, it serves as a powerful advocacy tool for securing funding for flood mitigation projects. 

However, 3D flood representation has remained with some challenges \citep{wang2019parallel}: (1) Fast model computation in which the flood modelling results should be provided within hours to serve various mitigation actions; (2) Effective information representation, in which visualization is the window to communicate flood risk to stakeholders in the community. To our best knowledge, the hydraulics software  (e.g., MIKE\footnote{\url{https://www.mikepoweredbydhi.com/products}}, Delft 3D\footnote{\url{https://oss.deltares.nl/web/delft3d}}) has undergone significant development to meet the target of decision-makers for accurate prediction of floods. However, they are more used by research-level organizations and hardly maximize their potential in communicating flood risks to stakeholders in the community.

In this context, this article proposes a 3D virtual geographic environment that integrates flood spatiotemporal modelling, parallel computation, and 3D representation like a workflow. Therefore, we aim to improve the computational efficiency of flood modelling and thus serve disaster management promptly. The 3D representation also allows flood modelling to move from specialized outputs to intuitive 3D scenes, providing a valuable platform for substantial risk communication among stakeholders, particularly for people without direct experience with floods who can perceive flood propagation from an all-round perspective.

The remainder of this article is organized as follows: In Section 2 the related work of this study is reviewed. Section 3.1 provides a study framework for this article. Sections 3.2, 3.3, and 3.4 introduce flood spatiotemporal modelling, parallel computation, and web 3D visualization, respectively. In Section 4 the case experiment is conducted and analyzed, and findings are summarized. Finally, Sections 5 and 6 present the discussion and conclusion.   
  \section{Related work}

The development of humankind is a history of struggle against disasters, and the exploration of disasters has gone through a long historical period and formed a series of mitigation strategies \citep{li2022investigations}. In 1936, the United States issued the Flood Control Act to enhance structural defences, and such mitigation measures gradually became the principal strategy in many countries \citep{samuels2006analysis}. Obviously, structural measures are the most direct but also the most expensive ways to mitigate flood hazards. Later on, some studies have shown that there is an illusion of perfect safety among populations protected by levees and dams, which makes the residents near these structures clearly ignore the flood risk and reluctance to evacuate \citep{spittal2005optimistic,meyer2012economic,kundzewicz2018flood,li2022investigations}. From the authors' perspective, we manage neither to keep destructive waters away from people at all times nor keep people away from destructive waters \citep{kundzewicz2018flood}. Aside from strengthening flood protection via structural ``hard" measures, improving the resilience of the flood system as a whole from a non-structural ``soft" management (e.g., flood warning, community awareness.) should be taken into account as well. For example, considering loss reduction and creating a resilient environment, the Geospatial Infrastructure Management Ecosystem (GeoIME) is used to measure and determine the vulnerability of buildings and risk estimation for disaster management \citep{pirasteh2020cloud}.

The European Commission (EC) has taken the lead in designing and progressively implementing non-structural technical solutions to reduce flood risk in Europe. For example, the EC launched the European Flood Alert System (EFAS) in 2002 to forecast and provide early warning information to its partner countries \citep{smith2016operational}. Subsequently, the European Union (EU) has dedicated legislation called Directive 2007/60/EC, which required the establishment of flood maps for high-risk areas in all member states by 2013 \citep{meyer2012economic,kundzewicz2018flood}. It was evident that geoinformatics have advanced in the last decade and opened a promising new path for flood risk management. 

Besides, flood mapping is a crucial component of risk management, which can help decision-makers improve preparedness and raise public risk awareness \citep{meyer2012economic,zhang2022mapping}. In general, flood mapping includes numerical modelling and map generation. Flood modelling utilizes a mathematical approach to solve equations that represent the physical behavior of floods \citep{hadimlioglu2020floodsim}. In general, there are one-dimensional (1D), two-dimensional (2D), and three-dimensional (3D) hydrodynamic models used to analyze and model floods. Compared with 1D models, 2D models are more widely used since an additional dimension can better describe the spread of floods, whereas 3D models require more parameter settings \citep{mudashiru2021flood}. Currently, there are many flood modelling tools exist for global users, such as MIKE, HEC-RAS, FLO-2D, and Delft 3D, which have undergone significant development to meet the target of decision-makers for predicting floods. However, these tools have complex parameter configurations, and the visualization lacks coupling to the geographic scene. The data exchange between modelling and visualization needs to involve a variety of third-party plug-ins or modules, which makes it difficult for non-hydraulic experts to use and hardly maximize their potential in communicating flood risks to stakeholders in the community.

Furthermore, map generation is a cartographic representation of flood information from a geospatial perspective, which helps simplify the complexity of flood events. As mentioned before, the typical products are paper maps and digital 2D maps, potentially valuable tools for promoting disaster risk understanding and communication \citep{costabile2021terrestrial}. Subsequently, some researchers gradually started to systematically discuss and introduce 3D visual representation into flood risk communication, which demonstrates a growing interest in supplementing flood maps with 3D visualization techniques to engage people with flood hazards \citep{lai2011development,yang2016gis,macchione2019moving}. Another trend for disseminating and visualizing 3D data is the application of web visualizations \citep{santis2018visual}, especially the development of WebGIS provides a new platform for the substantive exchange of flood information among stakeholders geographically distributed in different places, which has great potential to improve the efficiency of flood risk communication at the community level.

  \section{Methodology}
\subsection{Study framework}

\begin{figure}[h]
	\centering
	\includegraphics[width=0.85\textwidth]{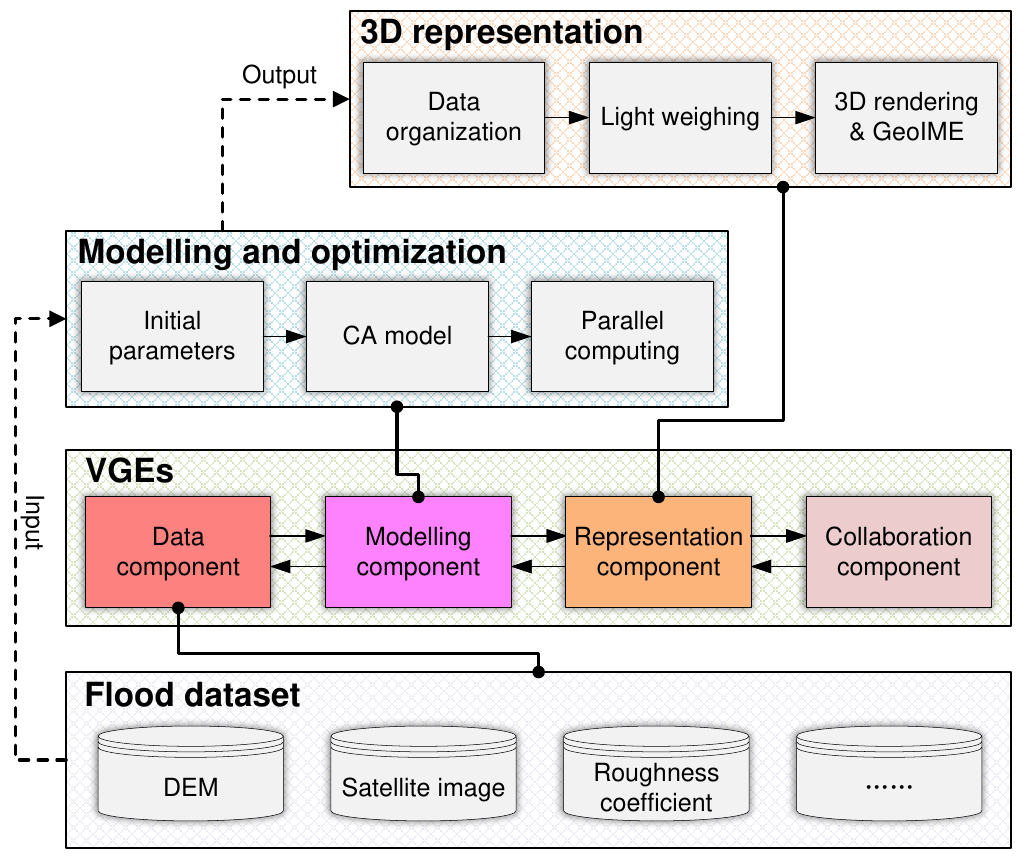}
	\caption{The proposed study framework of this article.} 
	\label{fig:fig1}
\end{figure}

The concept of virtual geographic environments was unofficially proposed for the first time in 1998, which can be described as a typical web- and computer-based digital geographic environment built for geographic understanding and problem-solving \citep{lin2013virtual,chen2018virtual,voinov2018virtual}. The core components of VGEs are data environment, modelling environment, representation environment, and collaboration environment. Following the connotation of virtual geographic environments \citep{lin2013virtual,lu2019reflections}, we design a virtual geographic environment-enabled integrated framework of flood 3D representation, as shown in \textcolor{red}{Fig.~\ref{fig:fig1}}. 

The flood dataset corresponds to the data environment, which is used to prepare the data required by flood spatiotemporal modelling and is the foundation for the framework construction. The modelling environment includes flood modelling and computation optimization. The flood geographic processes are modelled using the cellular automata (CA) model, and parallel computation is used to improve the solving efficiency. In the representation environment, we adopt WebGL-based 3D representation to visualize the spatiotemporal process of floods with the goal of improving risk communication.

\subsection{Spatiotemporal modelling of the flooding process}
\subsubsection{CA-based flood modelling}

\begin{figure}[tbh]
	\centering
	\includegraphics[width=0.75\textwidth]{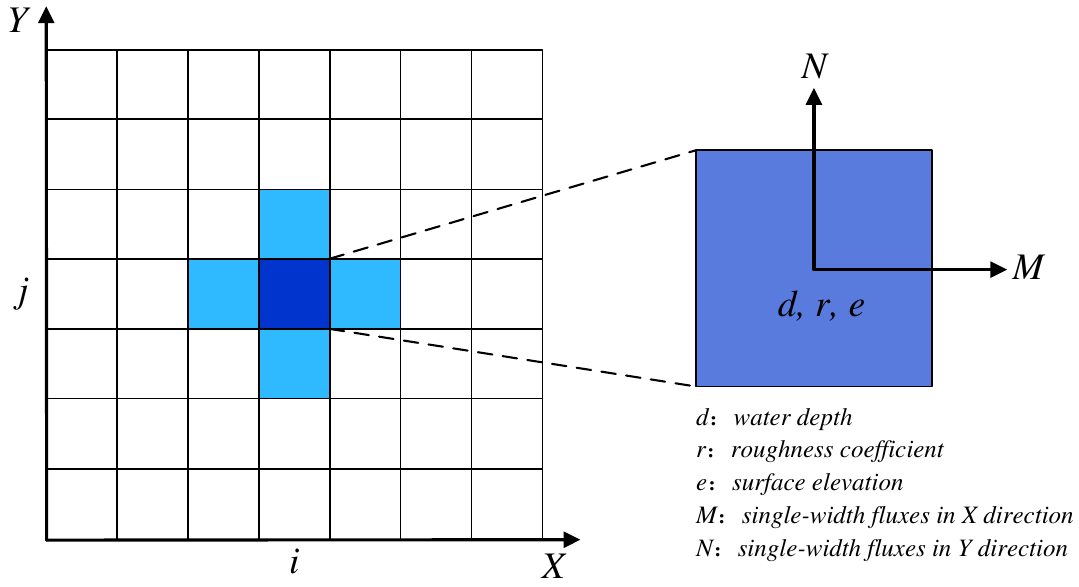}
	\caption{The structure of the CA-based numerical model of floods. The type of neighborhood is the Von Neumann neighborhood, which is composed of a central cell and four adjacent cells.}
	\label{fig:fig2}
\end{figure}

CA represents an alternative grid-based simulation framework, which divides geographic space into a series of discrete cells whose states will be updated by transition rules in discrete space-time, and has been increasingly used to simulate complex geographical phenomena \citep{li2013spatiotemporal,yeh2021cellular}, with the advantages of low data requirements, high computational efficiency, and suitability for parallel computing. In this context, we use a CA-based numerical model of floods \citep{li2013spatiotemporal,li2015real}, where each cell has five attributes, namely water depth $d$, roughness coefficient $r$, surface elevation $e$, single-width fluxes in X-direction $M$ and Y-direction $N$. \textcolor{red}{Fig.~\ref{fig:fig2}} shows the structure of the CA-based numerical model of floods.

The idea of value update of the $cell (i,j)$  is to solve Saint-Venant equations. The first step is to compute the single-width fluxes $M_{i, j}^{t+1}$ and $N_{i, j}^{t+1}$ at the moment $t+1$ based on the water depth $d_{i, j}^t$ and the corresponding neighbourhoods $d_{i+1, j}^t$, $d_{i, j+1}^t$ at the previous moment $t$, respectively, as shown in \textcolor{red}{Eq.~\ref{eq1}}.

\begin{equation}
\label{eq1}
\left\{\begin{array}{l}
M_{i, j}^{t+1}=M_{i, j}^t-g \frac{\Delta t\left(d_{i+1, j}^t+d_{i, j}^t\right)\left(z_{i+1, j}^t-z_{i, j}^t\right)}{\Delta x}-g r_{i, j}^2 \frac{\bar{u}_{i, j} \Delta t \sqrt{\left(u_{i, j}^t\right)^2+\left(v_{i, j}^t\right)^2}}{\left(\frac{d_{i+1, j}^t+d_{i, j}^t}{2}\right)^{\frac{1}{3}}} \\
N_{i, j}^{t+1}=N_{i, j}^t-g \frac{\Delta t\left(d_{i, j+1}^t+d_{i, j}^t\right)\left(z_{i, j+1}^t-z_{i, j}^t\right)}{\Delta y}-g r_{i, j}^2 \frac{\bar{v}_{i, j} \Delta t \sqrt{\left(u_{i, j}^t\right)^2+\left(v_{i, j}^t\right)^2}}{\left(\frac{d_{i, j+1}^t+d_{i, j}^t}{2}\right)^{\frac{1}{3}}}
\end{array}\right.
\end{equation}

where $d$ indicates water depth, $z$ is the sum of water depth $d$ and surface elevation $e$ names water stage, $r$ is roughness coefficient, $M$ and $N$ denote single-width fluxes in the $X$ and $Y$ directions respectively. $u$ and $v$ denote flow velocity in the $X$ and $Y$ directions, respectively. $\Delta t$ and the $\Delta x$, $\Delta y$ pair indicate the iteration unit time and cell size(m) respectively, $t$ and $t+1$ represent the current and the next moment, $g$ represents gravitational acceleration.

\begin{equation}
\label{eq2}
d_{i, j}^{t+1}=d_{i, j}^t-\frac{\Delta t\left(M_{i+1, j}^{t+1}-M_{i, j}^{t+1}\right)}{\Delta x}-\frac{\Delta t\left(N_{i, j+1}^{t+1}-N_{i, j}^{t+1}\right)}{\Delta y}
\end{equation}

As shown in \textcolor{red}{Eq.~\ref{eq2}}, the second step is to update the water depth $d_{i, j}^{t+1}$ at the moment t+1 based on the water depth $d_{i, j}^{t}$, the single-width fluxes $M_{i+1, j}^{t+1}$ and $M_{i, j}^{t+1}$ in the $X$ direction, and the single-width fluxes $N_{i, j+1}^{t+1}$ and $N_{i, j}^{t+1}$ in the $Y$ direction.


\subsubsection{Parameters configuration and computing workflow}

The water stage, the water depth, and single-width fluxes of inlet cells are generally initialized considering the historical flood data, the surface elevation values are extracted from DEM, the setting of Manning’s roughness coefficient with reference to \cite{li2013spatiotemporal}.

In addition, the following conditions need to be set to constrain the computing process of flood modelling. 

\begin{itemize}
\item The discharge is equal to the sum of cell water and the residual water;
\item The cell outside the boundary is regarded as an invalid cell, which is not involved in the iterative calculation;
\item The increase in single-width fluxes of a cell cannot exceed the fluxes it can provide;
\item If the number of neighborhood cells with water in a specific cell is more than 7, the cell changes to a water state, conversely, if the number of neighborhood cells with water in a specific cell is less than 2, the cell changes to a no-water state.

\end{itemize}

\begin{figure}[h]
	\centering
	\includegraphics[width=0.85\textwidth]{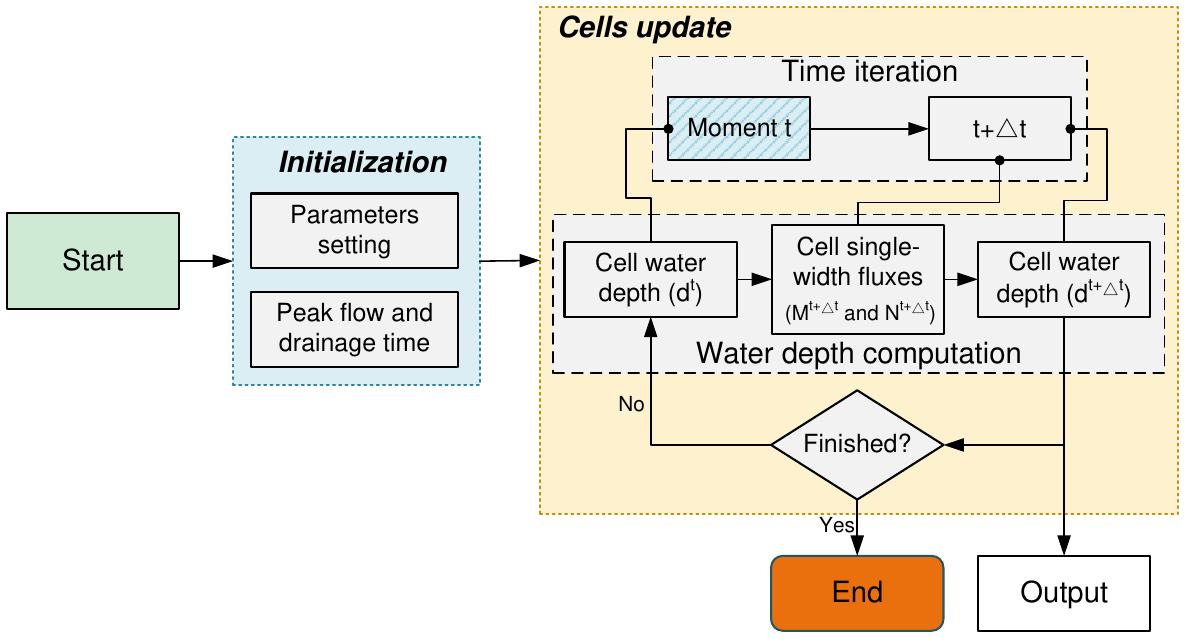}
	\caption{The computing workflow for the spatiotemporal modelling of the flood. }
	\label{fig:fig3}
\end{figure}

To give the reader a clearer understanding of the spatiotemporal modelling of the flood, we propose a computing workflow integrating the above constraints in \textcolor{red}{Fig.~\ref{fig:fig3}}. In the initialization part, some parameters (e.g., discharge, inlet cells) are set, the peak flow and drainage time are computed. In the cell update part, cell water depth $d^t$ at the current moment $t$ is used to compute the cell single-width fluxes $M^{t+\Delta t}$ and $N^{t+\Delta t}$ at the next moment $t+\Delta t$, then use this single-width fluxes to update the cell water depth $d^{t+\Delta t}$  at the moment $t+\Delta t$ and output the value of each moment. In the above loop, if the process ends then jump out of the loop, otherwise iterate the above process.

\subsection{Computation optimization based on Open Multi-Processing}
\subsubsection{Parallelism principles and evaluation indicators}

\begin{figure}[h]
	\centering
	\includegraphics[width=0.85\textwidth]{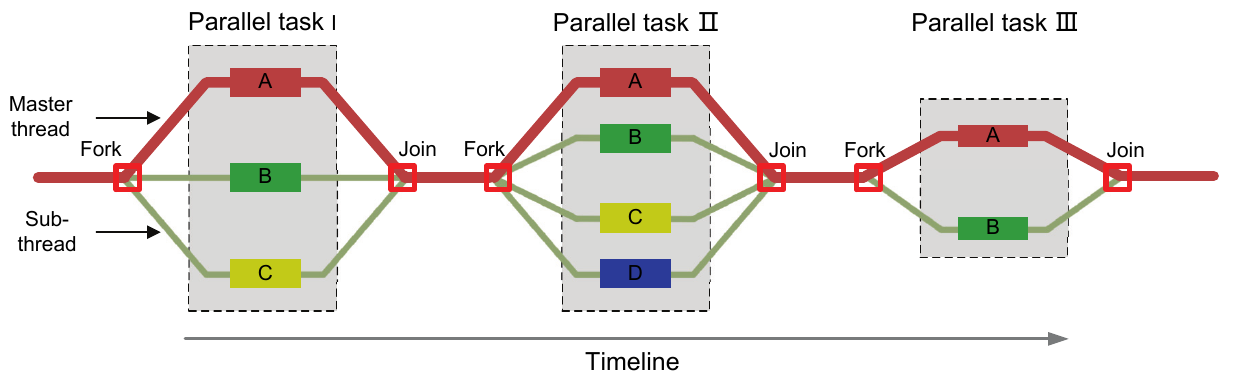}
	\caption{An illustration of the OpenMP parallel principles and execution process.}
	\label{fig:fig4}
\end{figure}

Open Multi-Processing (OpenMP) uses a fork-join parallel mode, where threads are opened and destroyed by forking/joining to achieve the purpose of multi-threaded parallelism \citep{aldinucci2021practical}, the principles are shown in \textcolor{red}{Fig.~\ref{fig:fig4}}. The master thread A executes at the start of the program, it will derive two sub-threads B and C when encounters a fork in the running. At this point, the three threads A, B, and C execute together, and this part is called the parallel task I. After the three threads have finished their computational tasks in the parallel zone, the sub-threads are joined and only the master thread A continues executing subsequent tasks until the next fork appears.

Furthermore, we use the speedup ratio $R_s$ to evaluate the computational efficiency of parallel algorithms, as shown in \textcolor{red}{Eq.~\ref{eq3}}.

\begin{equation}
\label{eq3}
R_s=\frac{t_s}{t_P}
\end{equation}

where $t_s$ indicates the time taken by a single CPU or core to complete the task. $t_p$ indicates the time taken by p CPUs or cores to complete the same task. 

\subsubsection{Dynamic parallelism considering load balance}

To enable parallel computation, designing the process and identifying independent tasks is a crucial step. Therefore, we analyzed the feasibility of parallel computation for flood modelling and designed a parallel scheme described as follows.

(1) Parallel scheme design

\begin{figure}[tbh]
	\centering
	\includegraphics[width=0.83\textwidth]{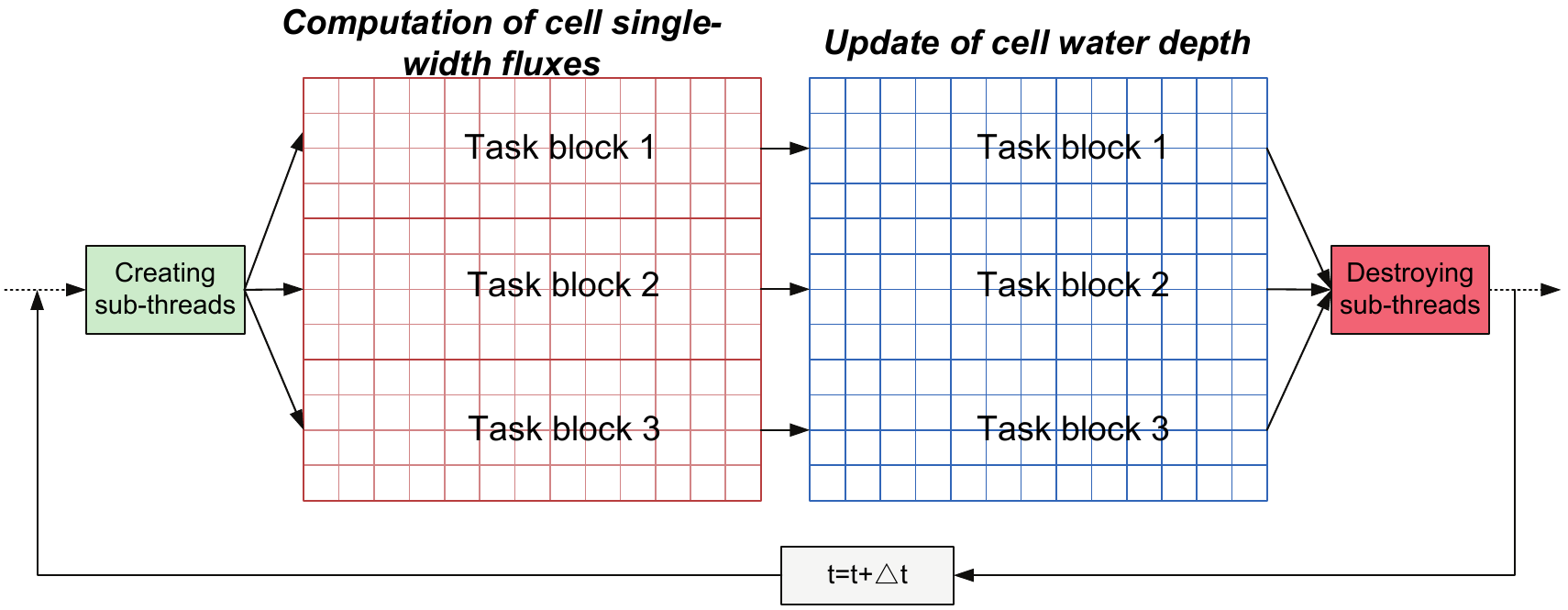}
	\caption{The parallel scheme for the computation of flood modelling.}
	\label{fig:fig5}
\end{figure}

The main effort to compute flood modelling concentrates on the state update of cells, which mainly involves single-width fluxes and water depth. For example, the computation of the single-width fluxes of the target cell at moment $t+1$ requires incorporating the water depth of the neighborhood cells at moment $t$. The single-width fluxes of the cell at the moment $t+1$ are used to update the water depth of the target cell at moment $t+1$. Still, these two parameters meet the independence of computation between different cells at the same time. Therefore, we create two computation tasks: the first is used to compute the single-width fluxes of cells, and the second is used to compute the water depth of cells, as shown in \textcolor{red}{Fig.~\ref{fig:fig5}}.

(2) Computational task division and scheduling

The task division is usually done by dividing the total of tasks by the number of threads and then obtaining the task size to be performed by each thread. In general, there are two types of scheduling for task execution: 

\begin{itemize}

\item Static scheduling: the amount of computation is divided into a fixed number of threads, and each thread is assigned to compute the corresponding task blocks, no further task scheduling is performed during the computation.
\item Dynamic scheduling: each thread is dynamically assigned the task block according to its computation speed, the thread with faster execution will get more task blocks, thus better utilizing the computation performance of each thread. 

\end{itemize}

In our case, as the flood propagates forward, the number of cells with water gradually increases, and the overall computation load becomes progressively larger, making the actual computation amount different for equal-sized task blocks. Therefore, dynamic scheduling is adopted to avoid the load imbalance caused by different computations in the parallel zone.

\begin{figure}[h]
	\centering
	\includegraphics[width=0.85\textwidth]{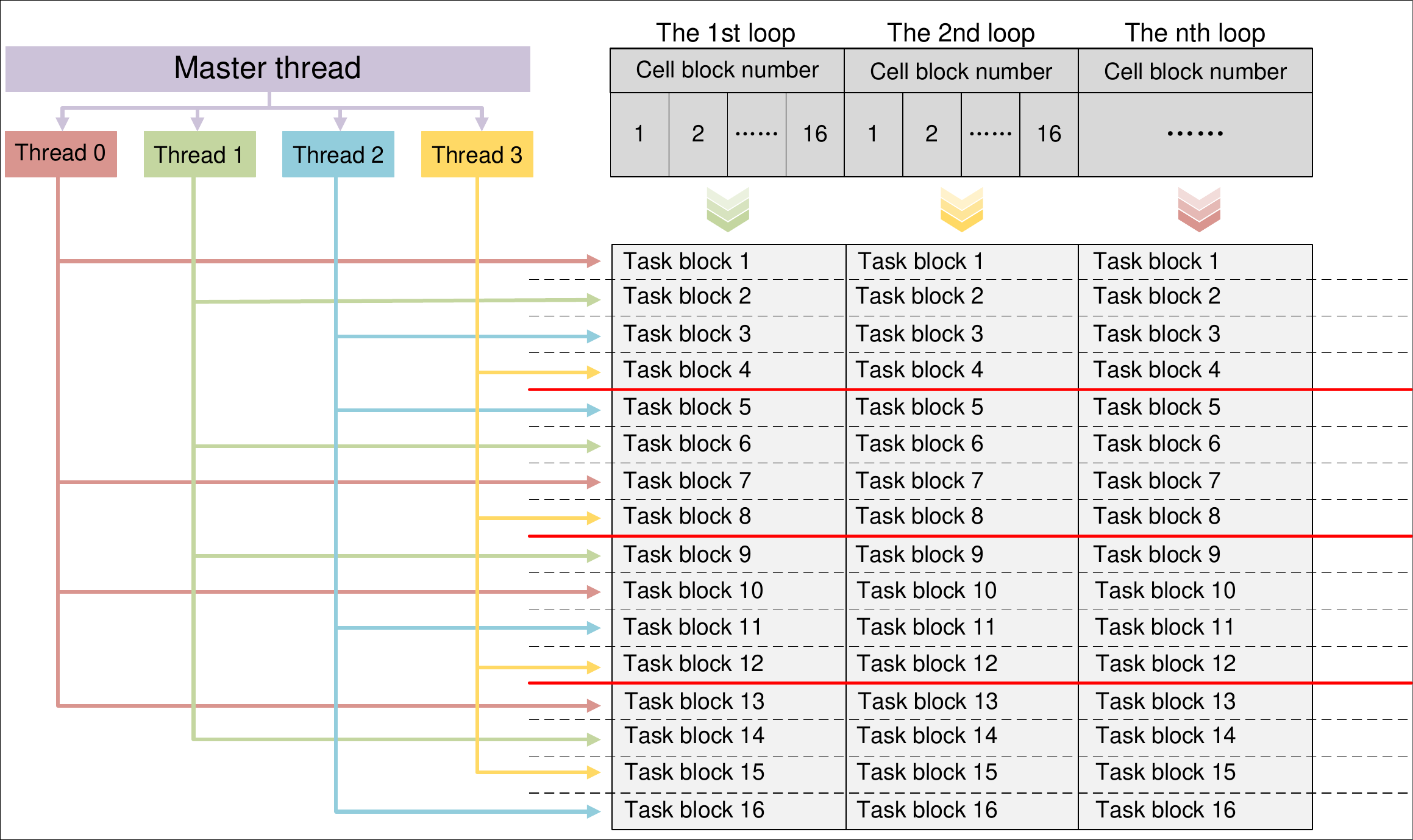}
	\caption{Computation task division and dynamic scheduling considering load balance.}
	\label{fig:fig6}
\end{figure}

Taking the computation of the single-width fluxes as an example, we divide the cell space into 16 blocks and use 4 threads to complete the computation task, as shown in \textcolor{red}{Fig.~\ref{fig:fig6}}. When the program starts execution, 4 threads sequentially read block 1 to block 4 from the queue, and the thread which finishes the current task block will continue executing the 5th task block. In this case, task block 5 will be executed by tread 2 which finishes the computation task first instead of thread 0.





\subsection{Three-dimensional representation of flood spatiotemporal process}
\subsubsection{Organization of flood numerical results}

\begin{figure}[h]
	\centering
	\includegraphics[width=0.87\textwidth]{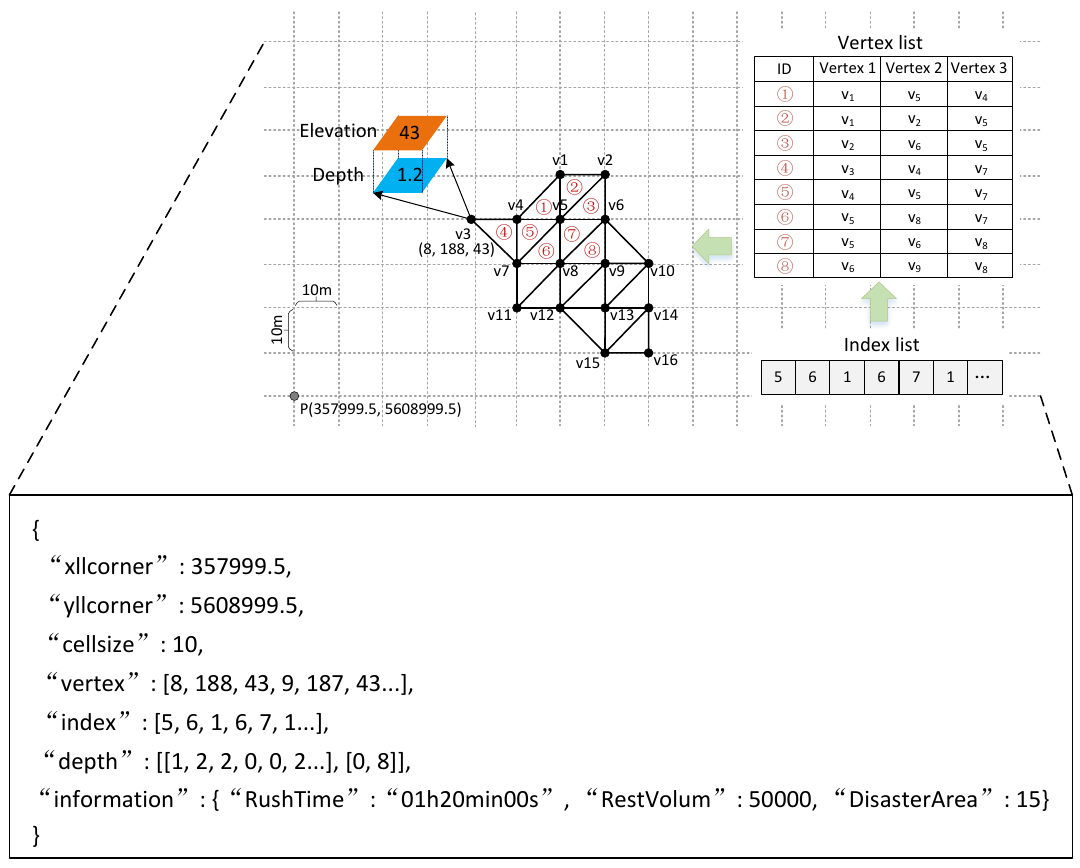}
	\caption{The data structure of flood numerical results.}
	\label{fig:fig9}
\end{figure}

The numerical modeling result of floods generally includes water depth, flow velocity, and coordinates of all cells, which can support flood mapping in 3D. Taking water depth as an example, we use JSON format to store and transfer the information mentioned above, and its data structure is shown in \textcolor{red}{Fig.~\ref{fig:fig9}}.

$P(xllcorner, yllcorner)$ is the reference point for converting the vertices from the local coordinate system to the global coordinate system. The cellsize indicates the spatial resolution of the flood surface. The vertex represents a list of vertices, starting from the $0^{th}$ element, and every three elements are marked as one vertex. The index indicates the index of the vertex. The depth is a 2D list, the first list includes the water depth values of all vertices and the second list includes the minimum and maximum values. The information indicates additional relevant information about flood modelling.

\subsubsection{WebGL-based flood 3D presentation to promote risk communication}

\begin{figure}[tbh]
	\centering
	\includegraphics[width=0.87\textwidth]{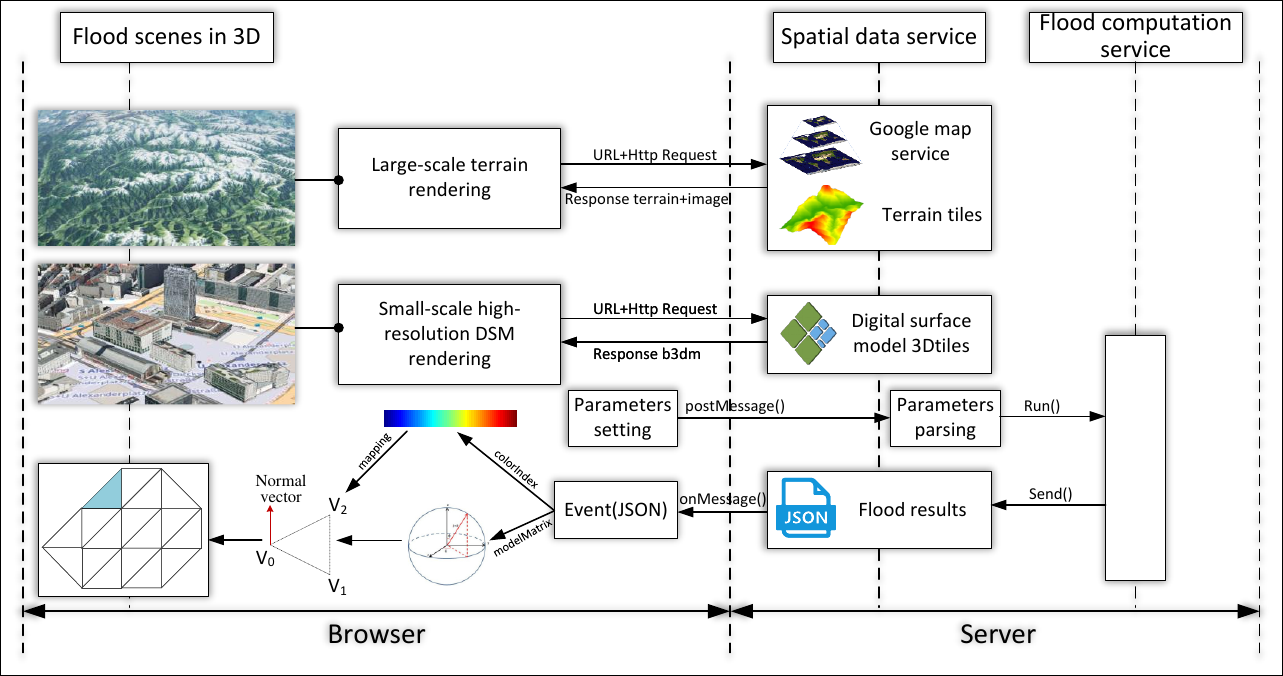}
	\caption{A plugin-free browser/server (B/S) framework for the flood representation in 3D.}
	\label{fig:fig10}
\end{figure}

\textcolor{red}{Fig.~\ref{fig:fig10}} shows a plugin-free browser/server (B/S) framework for the flood representation in 3D, which can be used on diverse devices (e.g., laptops, tablets, smartphones, etc.). The spatial data service provides two types of services: the overlay of DEM and satellite images for a large-scale terrain rendering, and the DSM 3D Tiles, which represent more detailed environmental information around floods. The parameters of flood modelling are set on the browser and passed to the server through postMessage() to start the flood computation service. The browser gets the results by listening to onMessage(). In the part of representation, the modelMetrix transforms the vertices in the local coordinate system into the world coordinate system and the colorIndex is mapped to the corresponding vertices, and finally, the flood surface will be generated by the WebGL Graphics Pipeline. 

  \section{Experimental analysis and results}
\subsection{Case area and data description}

\begin{figure}[h]
	\centering
	\includegraphics[width=0.87\textwidth]{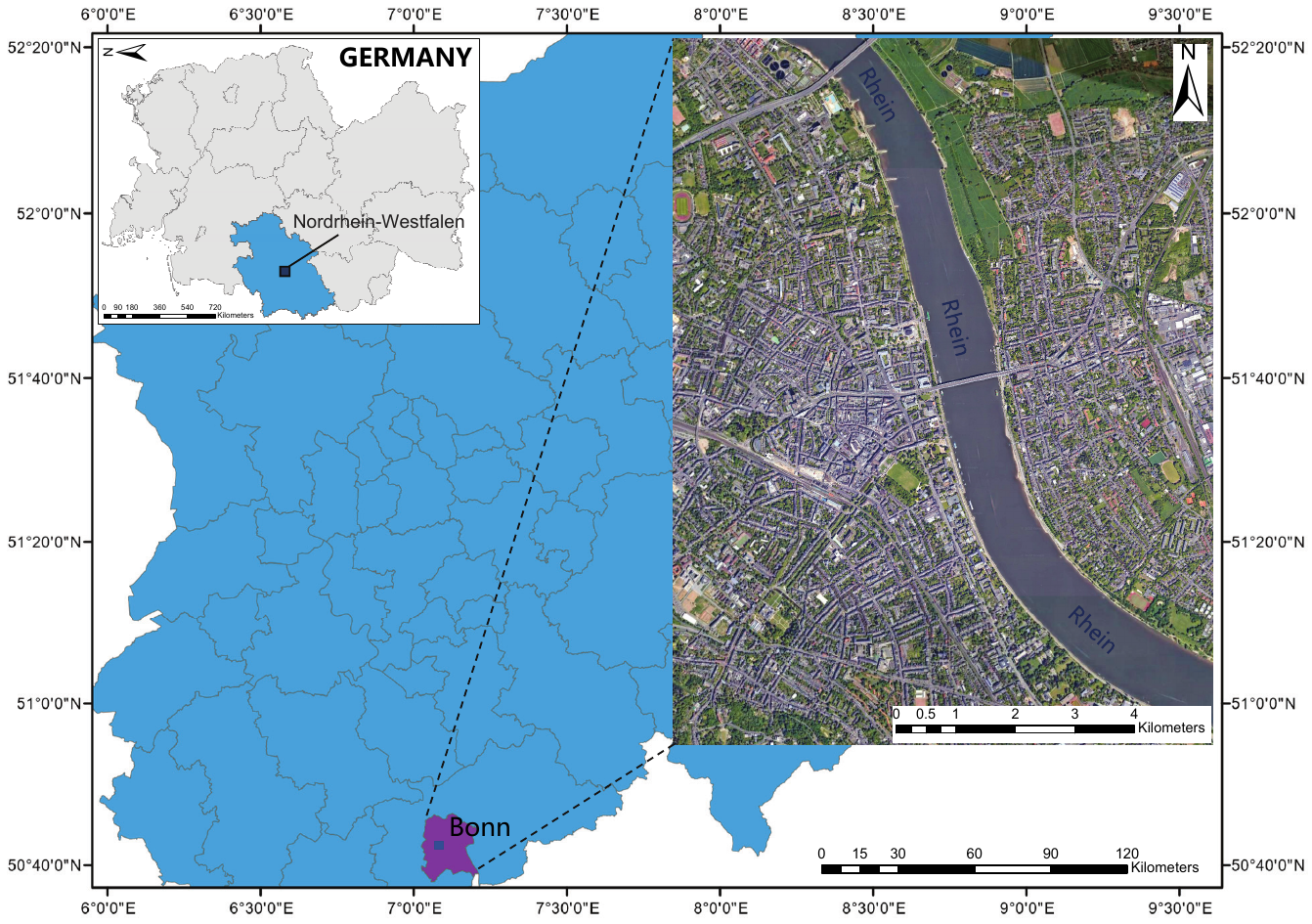}
	\caption{Study area: a section of the Rhine River in Bonn, Germany.}
	\label{fig:fig11}
\end{figure}

In this article, we selected a section of the Rhine River in Bonn, Germany, as the case area for experiment analysis, as shown in \textcolor{red}{Fig.~\ref{fig:fig11}}. The catastrophic flood that occurred in July 2021 hints at the threat of floods to Bonn due to global climate change, it also points out the need to promote flood risk communication and enhance the public risk awareness in this region.

We collected DEM (1m$\times$1m), buildings (polygon), roads (polyline), and DSM (0.5m$\times$0.5m). Specifically, DEM serves as the basis for flood modelling, while DSM is used for realistic landscape representation. In this context, buildings and roads are essential for subsequent risk assessment and also pave the way to address further aspects such as rescue and disaster management. The DEM, buildings, and roads were obtained from the geoinformation exchange platform in North Rhine-Westphalia (NRW)\footnote{\url{https://www.geoportal.nrw/}}, the DEM was further converted into a series of tiles, and the house and road data were converted into GeoJSON. The DSM was provided by the Office for Land Management and Geoinformation of the City of Bonn with nearly 50 gigabytes, which was processed into 3D Tiles and the corresponding index file was generated. The base maps are supported by OpenStreetMap (OSM)\footnote{\url{https://basemaps.cartocdn.com}} and ArcGIS Imagery\footnote{\url{https://server.arcgisonline.com}}.

\subsection{Experiment environment}

The experiment was implemented on Lenovo Legion R9000P2021H, whose processor is AMD Ryzen 7 5800H with Radeon Graphics, with 16 GB memory and NVIDIA GeForce RTX 3060 Laptop GPU 6 GB. The development of parallel optimization depends on the third-party library OpenMP 2.0. An open-source library, CesiumJS v1.73, was selected as the 3D rendering engine to represent the virtual landscape and flood information. NodeJS v16.18.0 and Google Chrome were used as the server and browser, respectively.

\subsection{Flood modelling results}

\begin{figure}[h]
	\centering
	\includegraphics[width=0.85\textwidth]{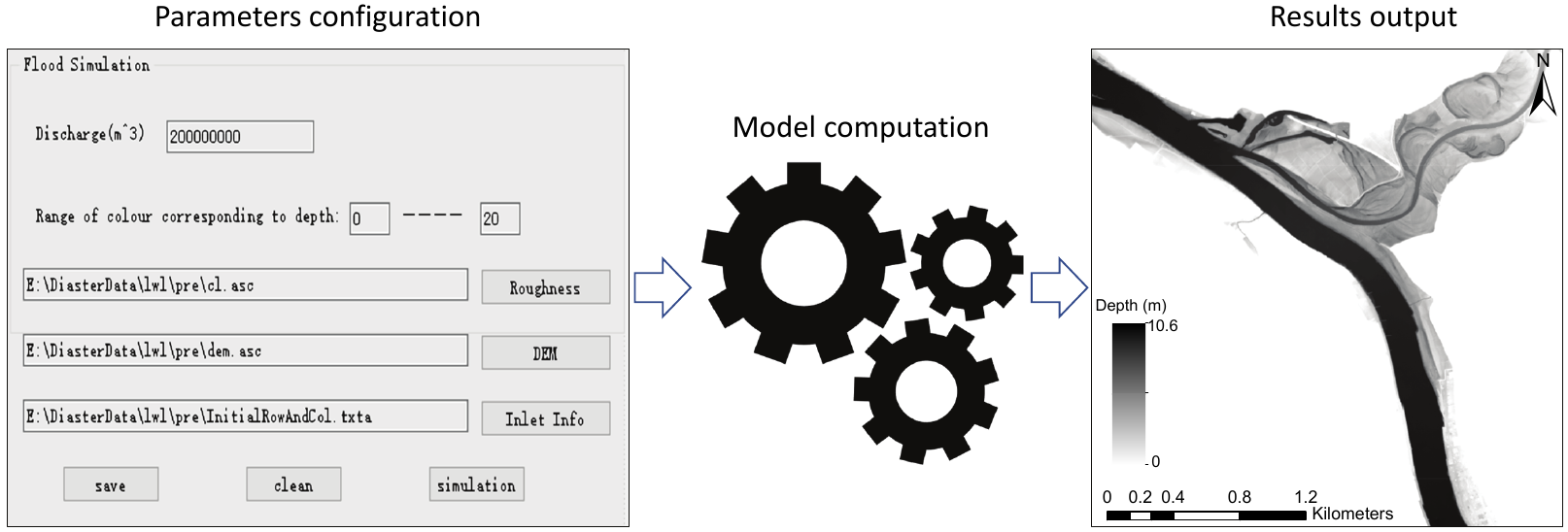}
	\caption{Parameters configuration and flood modelling results output.}
	\label{fig:fig12}
\end{figure}

First, we set the discharge value and colour mapping scheme. Second, the roughness coefficient and DEM required for the flood modelling were processed into ASCII format and their projection was set to ETRS\_1989\_UTM\_Zone\_32N. Finally, the row and column numbers and water depth values of the inlet cells were set. When the flood modelling process was finished, a time series of raster files were output, showing the inundation range and water depth of the flood's inundation range and water depth at different moments, as shown in \textcolor{red}{Fig.~\ref{fig:fig12}}.


\subsection{Efficiency analysis of computation optimization}

In general, serial computation can perform well in flood modelling for small areas, but it would be less computationally efficient for large-scale areas and high-resolution data input. Therefore, we used OpenMP-based parallel optimization to improve the computation efficiency of flood modelling, as mentioned in Section 3.3. The experiment of computation optimization in flood modelling was divided into two parts: static parallelism and dynamic parallelism. 

\subsubsection{Static parallelism}

\begin{figure}[h]
\centering
\subfigure[Time cost.]{%
\resizebox*{0.48\textwidth}{!}{\includegraphics{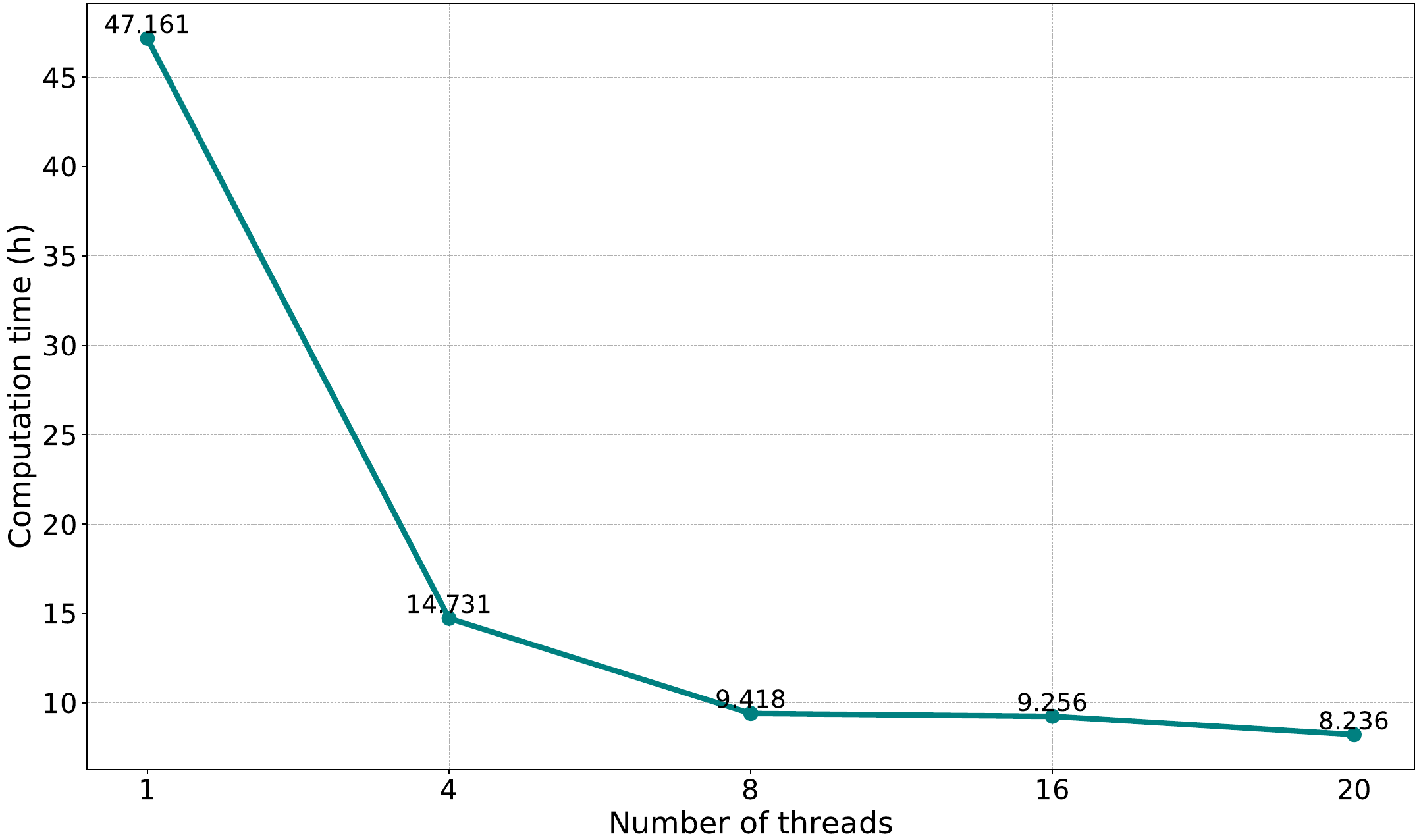}}}\hspace{3pt}
\subfigure[Speedup ratio.]{%
\resizebox*{0.47\textwidth}{!}{\includegraphics{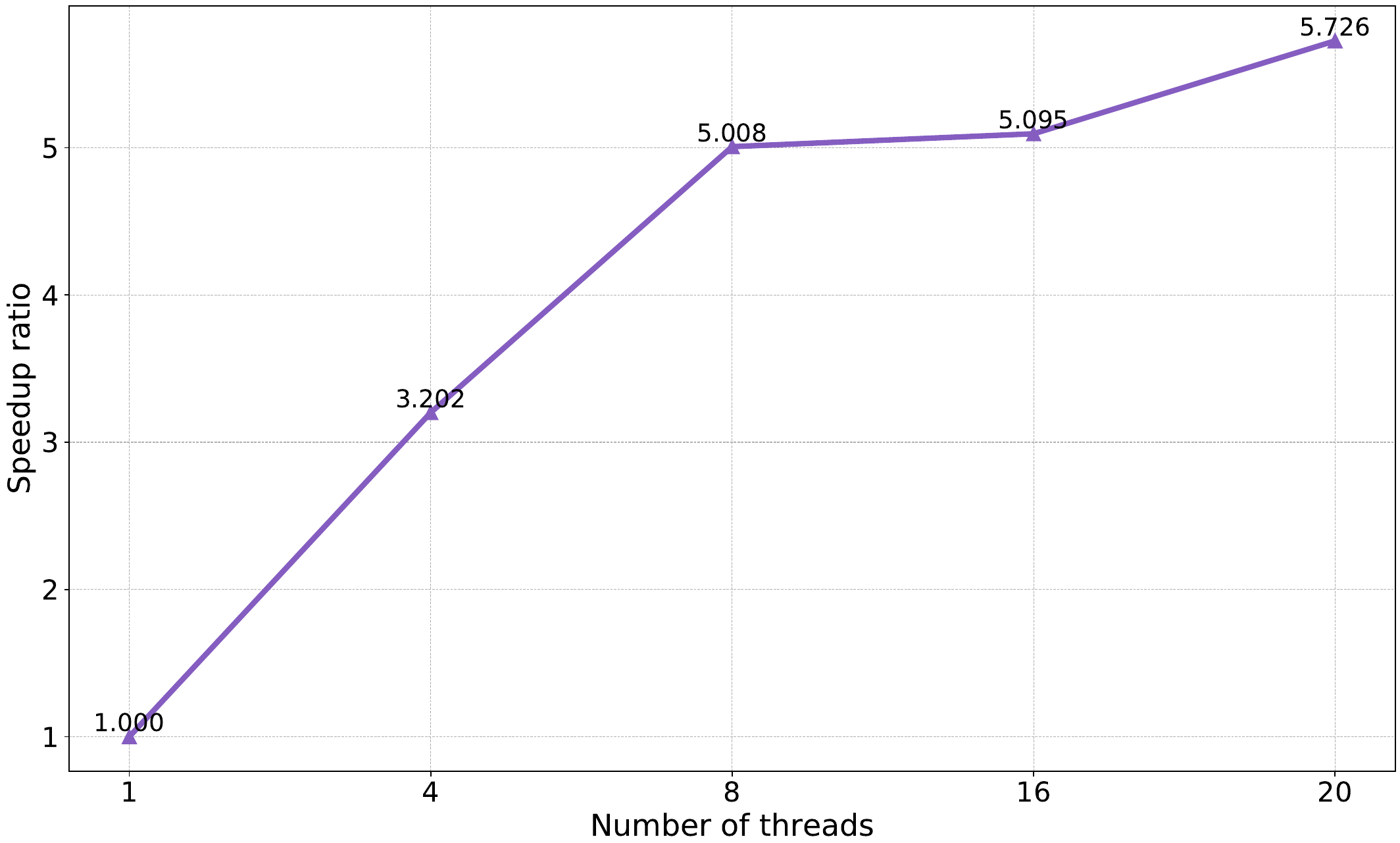}}}
\caption{Computation efficiency of static parallelism in flood modelling.} \label{fig:fig13}
\end{figure}

\textcolor{red}{Fig.~\ref{fig:fig13}} shows the computation efficiency of static parallelism in flood modelling, especially the time cost and speedup ratio trend with the increasing number of threads. As shown in \textcolor{red}{Fig.~\ref{fig:fig13}}(a), the time cost will shorten as the number of threads increases, the time cost is only 17.46\% of the serial computation (threads=1) when the threads were set to 20, and the speedup ratio increases as the number of threads increases accordingly (\textcolor{red}{Fig.~\ref{fig:fig13}}(b)). These two indices reveal that the increasing number of threads can indeed improve the computation efficiency in flood modelling.

Nevertheless, the increase in the speedup ratio will gradually slow down when the threads reach a certain number. For example, the speedup ratio increases from 1 to 5.008 when the number of threads increases from 1 to 8, while the number of threads increases from 8 to 20, and the speedup ratio increases only by 0.718. In general, a small number of threads is less expensive to schedule between threads, resulting in significant efficiency gains from parallel optimization at the beginning of the increase in the number of thread. Subsequently, the thread schedule will pay more cost as a large increase in the number of threads, thus making the speedup ratio increase slower.

\subsubsection{Dynamic parallelism}

\begin{figure}[h]
\centering
\subfigure[Time cost.]{%
\resizebox*{0.48\textwidth}{!}{\includegraphics{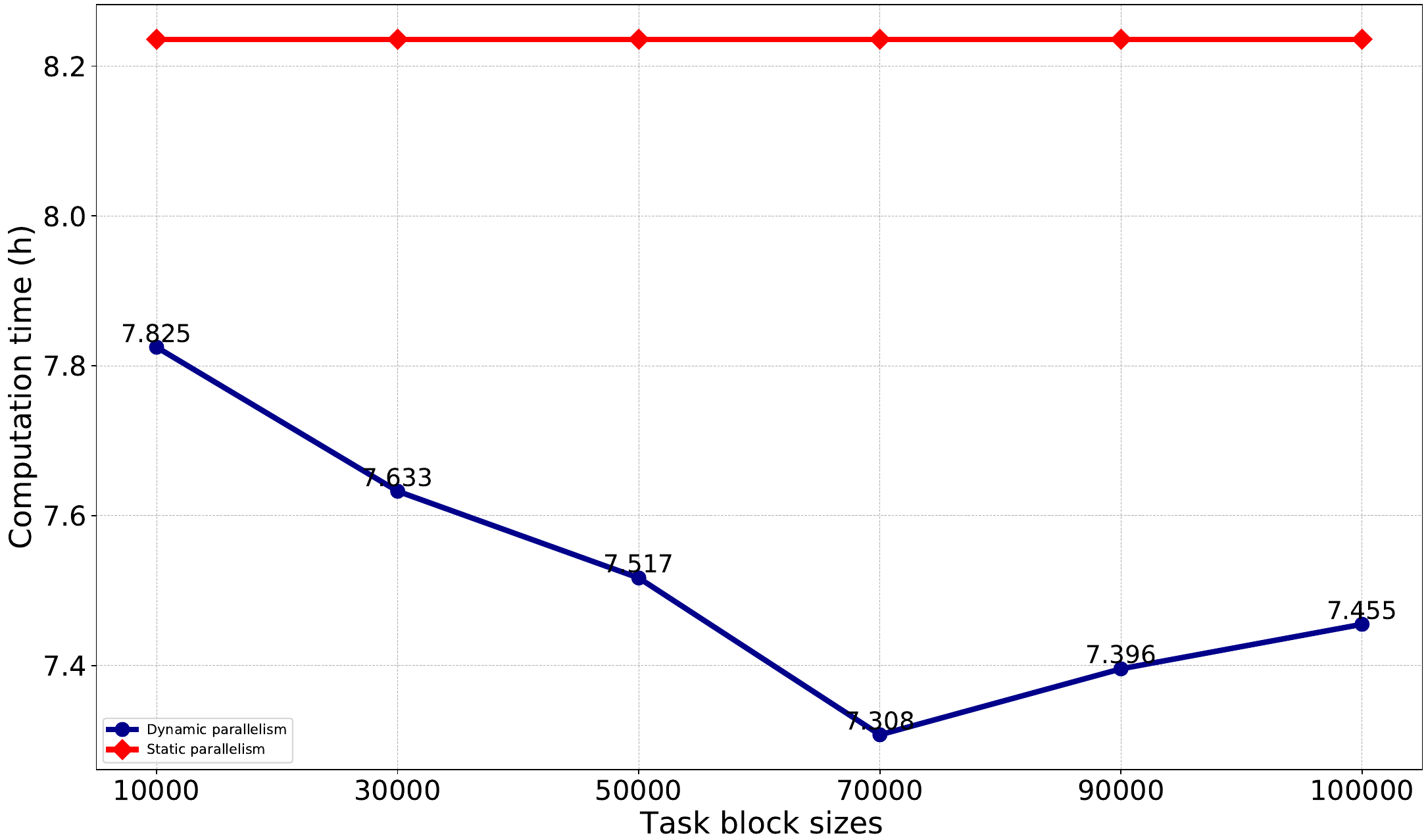}}}\hspace{3pt}
\subfigure[Speedup ratio.]{%
\resizebox*{0.48\textwidth}{!}{\includegraphics{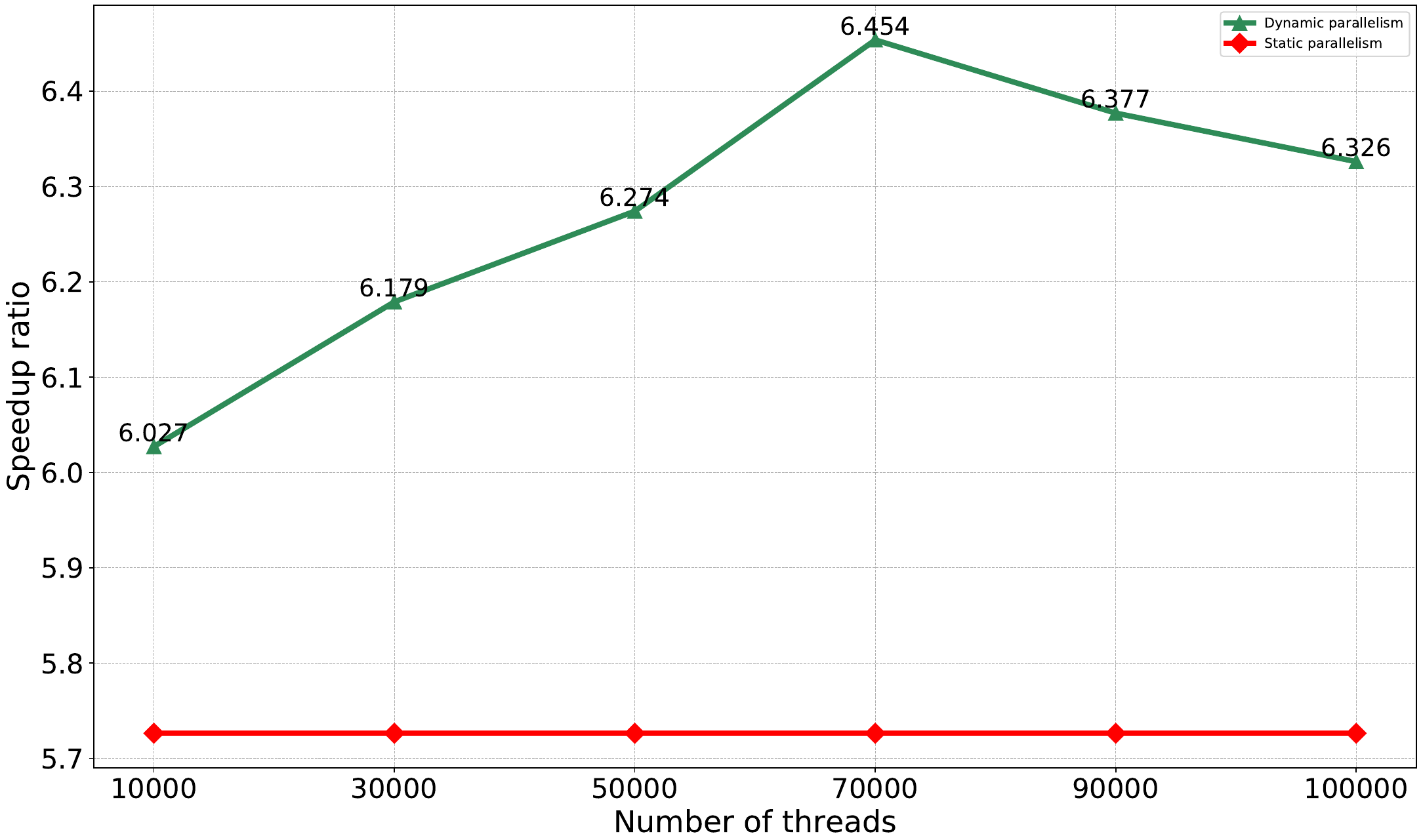}}}
\caption{Computation efficiency of dynamic parallelism in flood modelling.} \label{fig:fig14}
\end{figure}

We selected 20 as the fixed number of threads for the dynamic parallelism experiment and tested task blocks of sizes 10,000, 30,000, 50,000, 70,000, 90,000, and 100,000. The trend of time cost and speedup ratio with different task block sizes were shown in \textcolor{red}{Fig.~\ref{fig:fig14}}(a) and \textcolor{red}{Fig.~\ref{fig:fig14}}(b), respectively.

Dynamic parallelism with different task block sizes takes less time to compute than static parallelism under the same conditions, which indicates the former outperforms the latter in computation performance. However, it is not the case that the larger the task block size is, the better the computation performance. As the task block size increases, the time cost tends to decrease and then increase (\textcolor{red}{Fig.~\ref{fig:fig14}}(a)), and the speedup ratio tends to increase and then decrease (\textcolor{red}{Fig.~\ref{fig:fig14}}(b)). The speedup ratio peaks when the task block was set to 70,000, which indicates that 70,000 is the value close to the optimal block size in this experiment.

For dynamic parallelism with a fixed thread, a theoretical optimal block size achieves the maximum speedup ratio for that thread. On the one hand, this optimal size can reduce the waiting time for thread and avoid excessive schedule costs. On the other hand, it can also reduce the impact of load imbalance to some extent. Obviously, many comparative experiments are required to implement to get the optimal value.

\subsection{Flood mapping and 3D representation}

\begin{figure}[h]
	\centering
	\includegraphics[width=0.9\textwidth]{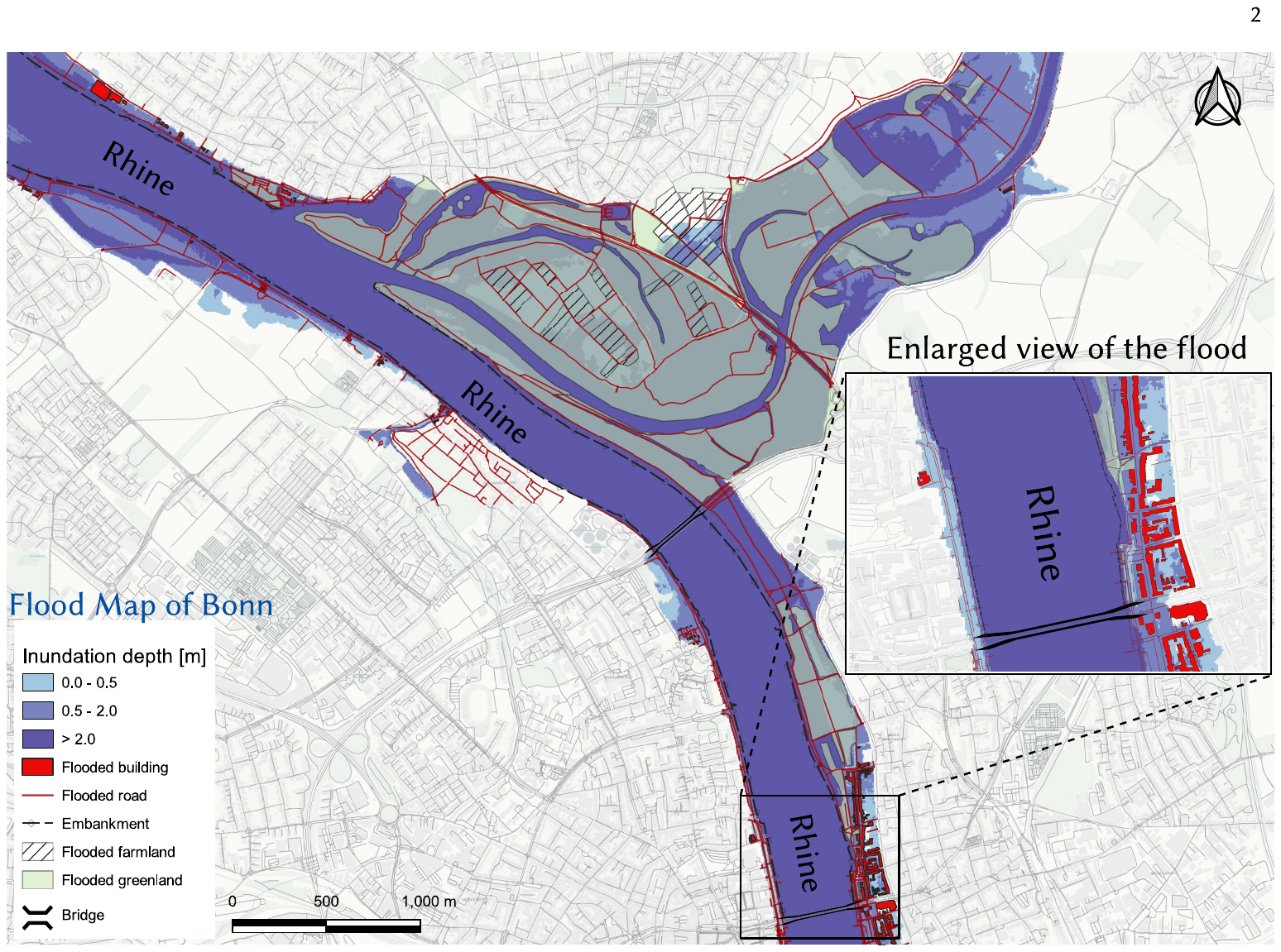}
	\caption{The flood map for parts of Bonn. We presented a section of the Rhine and its banks inundated by floods with the support of QGIS and OpenStreetMap. The mapping of inundation depth values and some cartographic rules were referenced by \cite{meyer2012recommendations}.}
	\label{fig:fig15}
\end{figure}

Building upon Section 4.3, we further produced a classic flood map of Bonn, as shown in \textcolor{red}{Fig.~\ref{fig:fig15}}. Map contents include inundation extent, flood depth, affected buildings, affected roads, key infrastructures, etc. Affected buildings and roads located within the inundation extent were highlighted on the map, which help users to determine evacuation routes and identify if their property would be affected, information that is also of primary interest to the general public \citep{meyer2012recommendations}. 

\begin{figure}[h]
\centering
\subfigure[A 2D view of ArcGIS maps.]{%
\resizebox*{0.49\textwidth}{!}{\includegraphics{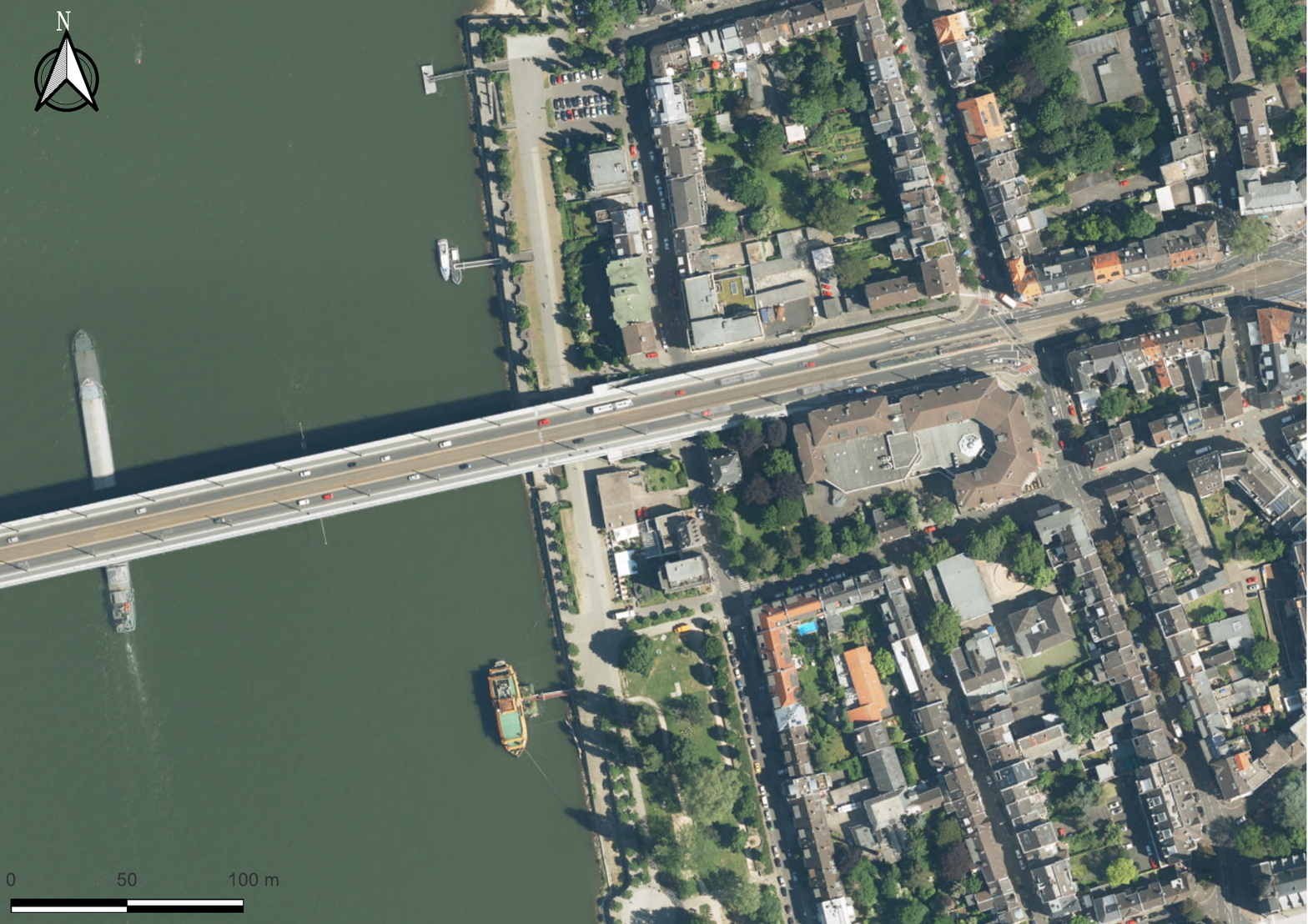}}}
\subfigure[A classic flood map based on satellite imagery.]{%
\resizebox*{0.49\textwidth}{!}{\includegraphics{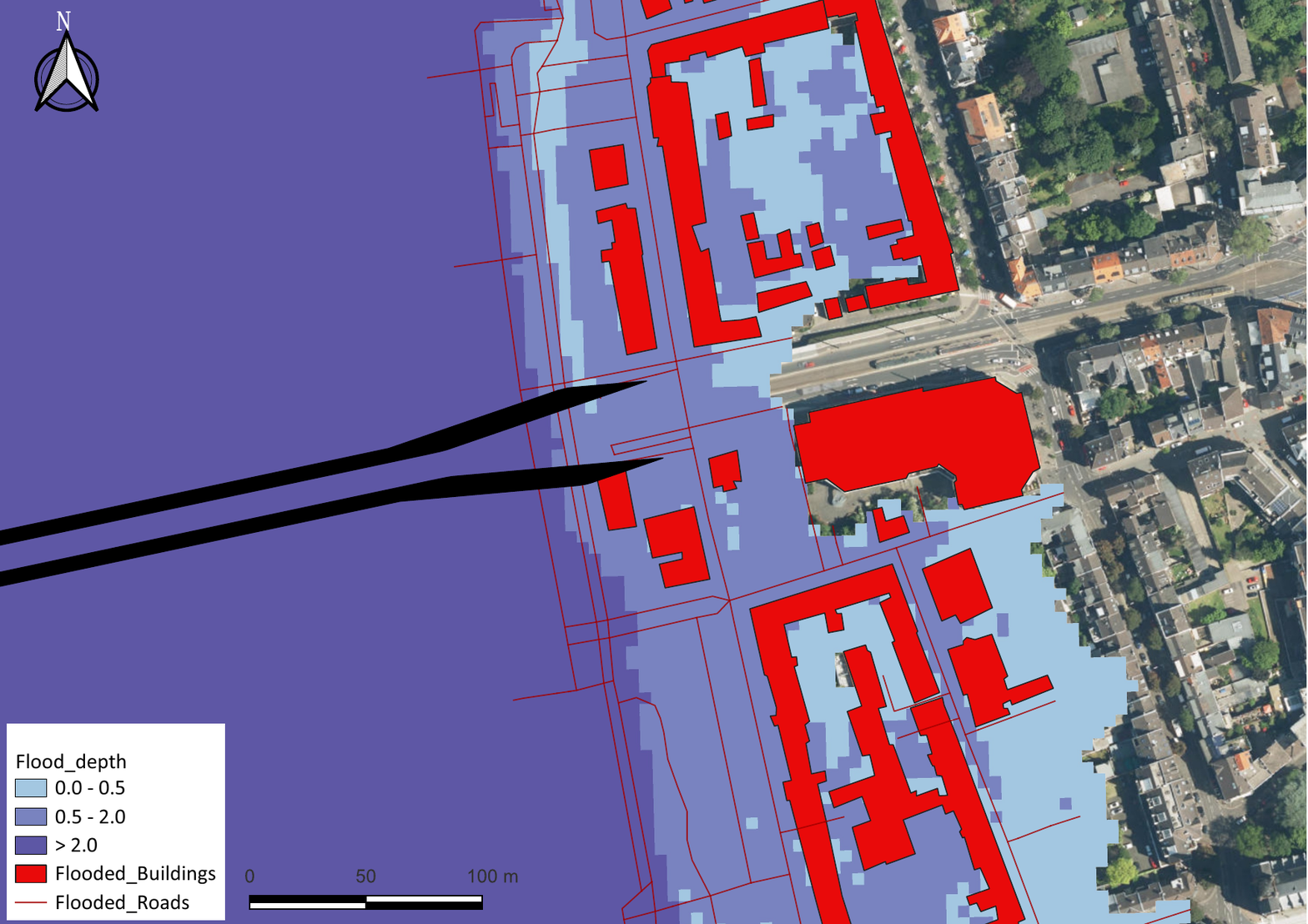}}}
\caption{Flood representation in a 2D view.} \label{fig:fig16}
\end{figure}

\begin{figure}[h]
\centering
\subfigure[A glance view of Bonn in 3D.]{%
\resizebox*{0.515\textwidth}{!}{\includegraphics{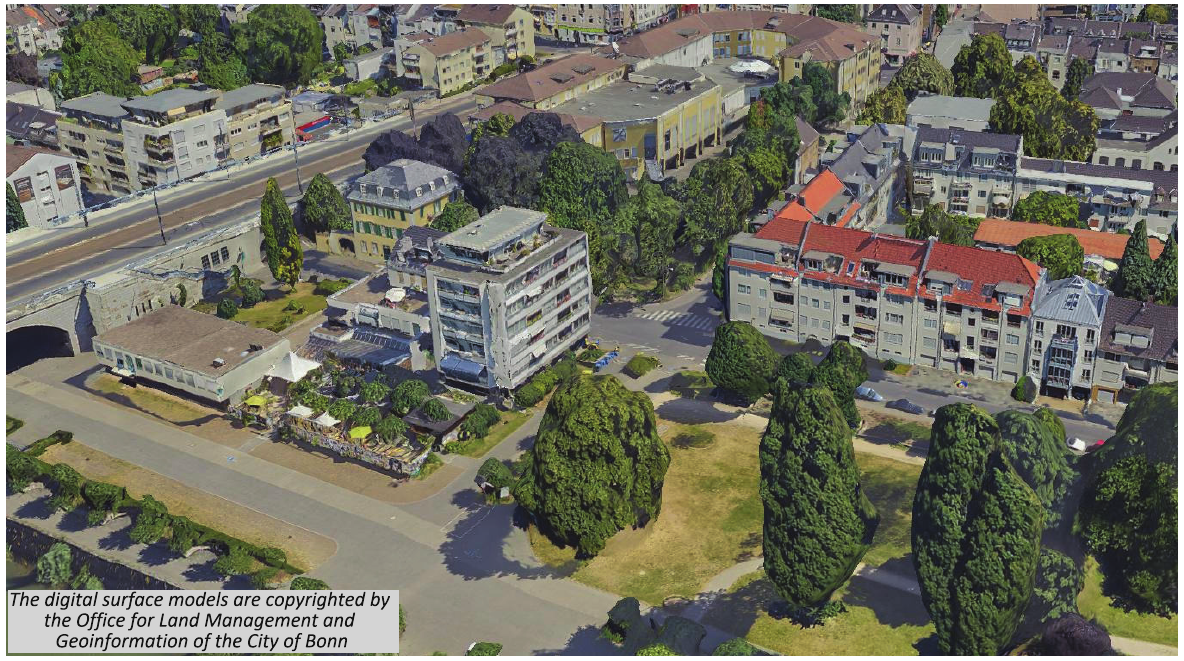}}}
\subfigure[Our flood representation in a 3D virtual environment.]{%
\resizebox*{0.47\textwidth}{!}{\includegraphics{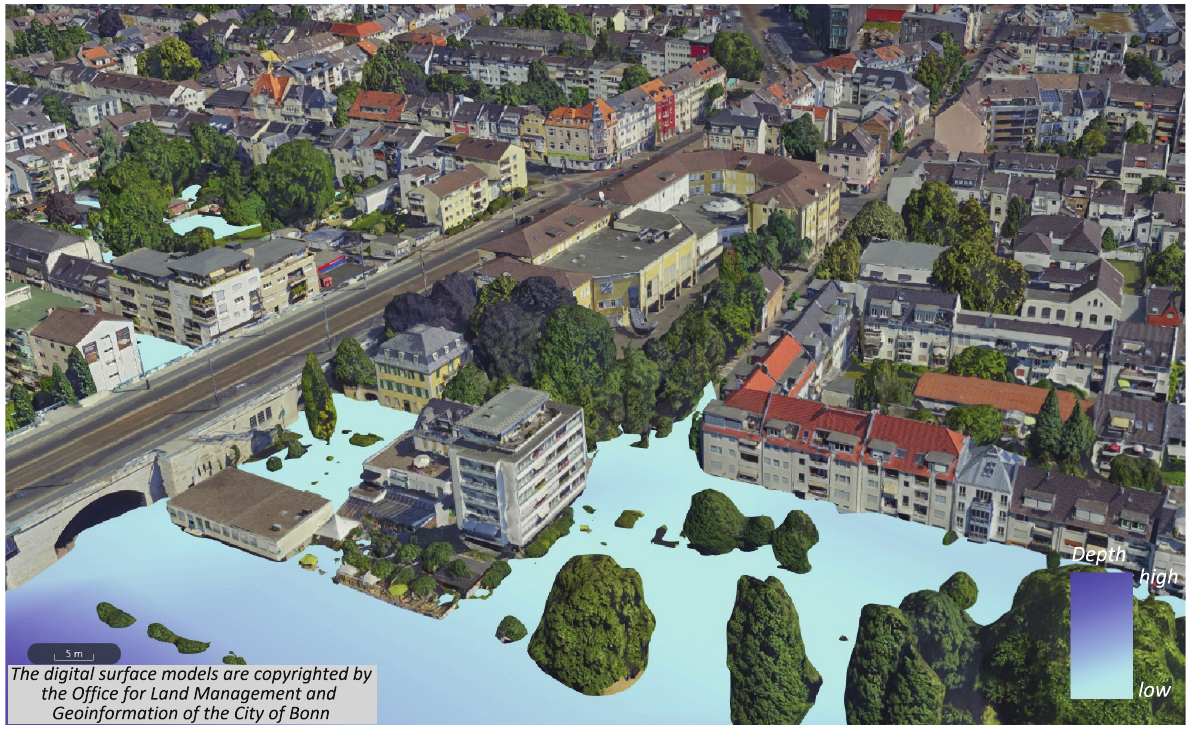}}}
\caption{Flood representation in a 3D view.} \label{fig:fig17}
\end{figure}

It is clear that flood maps play a vital role in strategic decisions on risk management and mitigation measures. However, due to the poor topographic auxiliary information and the non-intuitive visualization effect \citep{zhang2020efficient}, this kind of map makes it difficult for the general public without direct experience with floods to imagine what really happens. Therefore, we expanded the representation dimension of flood information and the corresponding landscape, from 2D to 3D, from abstract to realistic, which is beneficial for understanding the flooding process and promoting risk communication. \textcolor{red}{Fig.~\ref{fig:fig16}} and \textcolor{red}{Fig.~\ref{fig:fig17}} represent the flood information in 2D and 3D views, respectively. We documented the dynamic propagation process of floods in videos, which the readers can access at the following websites\footnote{\url{https://youtu.be/VWcD9AA1EfY}}$^{, }$\footnote{\url{https://youtu.be/j56422qF_D4}}.

  \section{Discussion}



\cite{li2021rapid} mentioned that parallel computation and web visualization as top priorities for future upgrades of flood modelling, thus this article is an extension and deepening of the work. Taking the theory of virtual geographic environment as a basis, a novel framework integrating flood modelling, parallel computation, and web visualization in 3D was proposed, which can accelerate the process of flood modelling and enhance more effective information representation, thereby helping people without direct experience with floods to better understand the flooding process and improving risk communication at the community level.

Specifically, the 3D virtual geographic environment-enabled integrated framework reduced the complexity and threshold of flood modelling and representation. As stated by \cite{kulkarni2014web} and \cite{mudashiru2021flood}, flood modelling mostly serves research-level organizations and is difficult for stakeholders to use due to the complexity of hydrologic models, cumbersome parameter settings, and professional mapping. In this article, we have simplified the parameter configuration and used a more intuitive 3D representation, which is more accessible to use for non-hydraulics professionals. 

The uncertainty of disasters requires a real-time or near-real-time emergency response \citep{voigt2016global,hu2018construction}. Therefore, a key challenge is the rapid computation of flood numerical models, which plays an important role in risk management by triggering the reduction of impact strategies \citep{wang2019parallel}. We adopt OpenMP-based parallel optimization to improve the efficiency of flood modelling, the dynamic parallelism considering load balance was used for the cell state update, which saves the time cost of thread waiting and fully exploits the usage of each thread. 

\cite{macchione2019moving} pointed out that flood maps are not characterized by a good balance between simplicity and complexity with adequate readability and usability for the public. In our case, the flood information representation was transformed from 2D to 3D, from abstract to realistic. On the one hand, it allows flood modelling to move away from specialized outputs to intuitive 3D scenes. On the other hand, it is helpful for participants from different backgrounds to understand the flooding process efficiently. In addition, \cite{santis2018visual} mentioned another trend for flood information dissemination is the application of web 3D visualization. The WebGL-based 3D representation that we used in this article can support the stakeholders distributed in different places to perceive the flood information, and it also supports tablets and smartphones very well, which is of great benefit to promoting flood risk communication in the community.




  \section{Conclusion}

With the support of geoinformatics, location-based risk communication to stakeholders can provide insight and decision support in risk management. In this article, we proposed a 3D virtual geographic environment for flood representation to promote risk communication. We first adopt CA-based numerical modelling to simulate the spatiotemporal process of floods. Then, parallel optimization is used to accelerate the computation efficiency of flood modelling. In addition, WebGL-based 3D representation is used to visualize the flooding process and its surrounding detailed environmental landscape. The application of the developed framework has been demonstrated for a section of the Rhine in Bonn, Germany. The results from the case study show that the framework can support the fast modelling and 3D representation of floods, which can be used as an effective tool for participants from different backgrounds to understand the flood process efficiently, thus promoting flood risk communication and disaster perception among various stakeholders.

In the future, we have two main tasks that need to be addressed as a priority, besides the limitations of this article proposed in the discussion section. First, spatial cognition and eye tracking will be introduced into the evaluation of flood risk communication. Second, exploring a more efficient interaction of flood scenes with the support of immersive virtual reality. However, this work has a great potential to be embedded on the GeoIME cloud-based application in future.

 \section*{Acknowledgements}


The authors would like to express their gratitude to the Office for Land Management and Geoinformation of the City of Bonn for the high-resolution digital surface model data and the geoinformation exchange platform in North Rhine-Westphali (NRW) for the thematic data, as well as the editor and anonymous reviewers for their careful reading and constructive suggestions.



\section*{Disclosure statement}

No potential conﬂict of interest was reported by the author(s).

\section*{Funding}

This paper was supported by the National Natural Science Foundation of China (Grant Nos. U2034202, 42201446 and 42271424), Postdoctoral Fellowship Program of CPSF (GZC20232185), Guangdong Science and Technology Strategic Innovation Fund (the Guangdong–Hong Kong-Macau Joint Laboratory
Program, Grant No. 2020B1212030009).



\bibliography{Reference}
\bibliographystyle{plainnat}

\end{document}